\documentclass[a4paper,english,aps,prb,twocolumn,floatfix,showpacs,amsfonts,amssymb,superscriptaddress]{revtex4}
\usepackage{graphics}
\usepackage{amssymb}
\usepackage{amsmath}
\usepackage{overpic}

\newcommand{\be}{\begin{equation}}
\newcommand{\ee}{\end{equation}}
\newcommand{\bea}{\begin{eqnarray}}
\newcommand{\eea}{\end{eqnarray}}

\def\a{\alpha}

\def\e{\varepsilon}
\def\d{\delta}

\def\m{\mu}
\def\l{\lambda}

\def\o{\omega}

\def\s{\sigma}

\def\D{\Delta}

\def\ra{\rightarrow}

\def\pd{\partial}

\def\nn{\nonumber}
\def\lb{\label}
\def\pref#1{(\ref{#1})}

\newcount\bozza \bozza=0
\ifnum\bozza=1
\newdimen\shift \shift=-2truecm
\def\lb#1{%
{\label{#1}\rlap{\kern\shift{$\scriptstyle#1$}}}}
\else\def\lb#1{\label{#1}} \fi

\begin{document}

\title{Doping dependence of the vortex-core energy in bilayer films of
  cuprates}

\author{L.~Benfatto}
\affiliation
{Centro Studi e Ricerche ``Enrico Fermi'', via Panisperna 89/A, I-00184,
  Rome, Italy}
\affiliation
{CNR-SMC-INFM and Department of Physics, University of Rome ``La
  Sapienza'', P.le Aldo Moro 5, 00185, Rome, Italy}

\author{C.~Castellani}
\affiliation
{CNR-SMC-INFM and Department of Physics, University of Rome ``La
  Sapienza'', P.le Aldo Moro 5, 00185, Rome, Italy}

\author{T.~Giamarchi}
\affiliation{DPMC- MaNEP University of Geneva, 24 Quai Ernest-Ansermet CH-1211
Gen\`eve 4, Switzerland}

\date{\today}

\begin{abstract}

The energy needed to create a vortex core is the basic ingredient to
address the physics of thermal vortex fluctuations in underdoped
cuprates. Here we theoretically investigate its role in the occurrence of
the Beresinskii-Kosterlitz-Thouless transition in a bilayer film with
inhomogeneity.  From the comparison with recent measurements of the
penetration depth in two-unit cell thin films we can extract the value
of the vortex-core energy $\mu$, and show that $\mu$ scales linearly with
$T_c$ at low doping.

\end{abstract}


\maketitle

One of the most puzzling aspects in the physics of high-temperature
superconductors (HTSC), that makes them substantially different from
conventional superconductors, is the separation between the fundamental
energy scales associated to superconductivity: the critical temperature
$T_c$, the zero-temperature superconducting (SC) gap $\D$ and the
superfluid stiffness $J_s=\hbar^2 \rho_s d_\perp/4m$.\cite{review_lee} Here
$\rho_s$ is the superfluid density, measured trough the London penetration
depth $\l$, and $\rho_s d_\perp$ is an effective two-dimensional superfluid
density, with $d_\perp$ a characteristic transverse length scale (see
discussion below).  While in a BCS superconductor the gap formation
and the appearance of superfluid currents
happen simultaneously at $T_c$, with $\D\sim T_c$, in the HTSC at low
doping level the two phenomena are essentially decoupled, and $T_c\sim
J_s$. This suggests that the transition can be controlled by phase
fluctuations, described within an effective $XY$ model for the phase
degrees of freedom, where $J_s$ sets the scale of the phase coupling. On a
general ground, also the energetic cost $\mu$ needed to
create the vortex core  is connected to $\lambda^{-2}$, i.e to $J_s$.
Using standard BCS relations one can see that at
$T=0$ both $J_s$ and $\mu$ are of order of the Fermi energy, which is a
large energy scale compared to $T_c$. Nonetheless, in thin films of
conventional superconductors $J_s(T)$ goes to zero as $T$ approaches
$T_{BCS}$, and $\mu(T)\sim J_s(T)$ is reduced, so that vortex creation
becomes possible but only at a $T_c$ very near to $T_{BCS}$.

In underdoped cuprate superconductors, where the BCS picture fails, a clear
 understanding of the typical energy scale which controls the vortex-core
 formation is still lacking. In particular, despite the fact that several
 experiments\cite{ong_nerst_prb06,ong_natphys07} suggest a predominant role
 of vortex fluctuations,\cite{tesanovic}
 whose occurrence is controlled in a crucial way by the value of
 $\mu$,\cite{review_minnaghen,benfatto_kt_rhos,benfatto_kt_magnetic,raghu}
 not much attention has been devoted yet to characterize the vortex-core
 energy $\mu$, and its relation with $T_c$.  In this paper we propose a
 procedure to estimate $\mu$ using penetration-depth measurements in thin
 films.
%Despite the intense activity that has been devoted to establish a clear
%relation between $J_s$ and $T_c$ in
%cuprates\cite{martinoli_films,triscone_fieldeffect,lemberger_films},
Indeed, in this case the transition is ultimately of
Beresinkii-Kosterlitz-Thouless (BKT) type, as it is signaled by the
linear relation between $T_c$ and $J_s$, and its persistence as doping is
changed by electric-field effect\cite{martinoli_films,triscone_fieldeffect}
or chemical doping.\cite{lemberger_films} To clarify the role played by the
vortex-core energy, let us recall some basic features of the BKT physics in
a SC film made of few unit cells. In Y-based cuprates, whose data we will
analyze below, there are two strongly-coupled CuO$_2$ layers within each
cell, which will be considered in what follows as the basic 2D unit of the
layered system, of thickness $d=12$ \AA, corresponding to the unit-cell
size in the $c$ direction. Then, for a $n$-cells thick SC film one would
expect the system to behave as an effective 2D superconductor with areal
density $\rho_s^{2d}=nd\rho_s$, where $\rho_s$ is the 3D superfluid density
connected to the penetration depth $\l$. The energy scale to be compared to
the temperature is then
\be
\lb{defj}
J_n=\frac{\hbar^2\rho_s^{2d}}{4m}=
\frac{nd\hbar^2\rho_s}{4m}=\frac{nd}{\l^2}\frac{\Phi_0^2}{4\pi^2\m_0}
\ee
where $\Phi_0=(h/2e)$ is the flux quantum, $\mu_0$ the vacuum permittivity
and we used MKS units as in Ref.\ [\onlinecite{lemberger_films}]. According
to BKT theory, a transition is expected when $J_n/T$ equals the universal
value $2/\pi$, which gives in terms of $1/\l^2(T)$ the relation:
\be
\lb{jump}
0.62 \times \frac{n d[{\mathrm{\AA}}]}{\l^2(T_{BKT}^n)[\m m^2]}=\frac{2}{\pi}T_{BKT}^n[K].
\ee
This approach assumes that all the layers of the film are so strongly
coupled that no mismatch of the SC phase in neighboring layers is possible,
i.e. the film behaves as an effective 2D layer of total thickness
$nd$. However, in the case of cuprates it is well known that the inter-cell
Josephson coupling $J_\perp$ is very weak, so that except in a narrow
region around $T_c$ one expects to see the BKT behavior of a {\em single}
unit cell. As far as the superfluid-density behavior is concerned, one
would then expect that the superfluid density drops already when Eq.\
\pref{jump} with $n=1$ is satisfied, which is certainly the case for
totally uncoupled cells. The best situation to analyze this effect in real
systems is provided by the finite-frequency sheet conductivity measurements
of thin Y$_{1-x}$Ca$_{x}$Ba$_{2}$Cu$_{3}$O$_{7-\d}$ (YBCO) films by Hetel et
al.\cite{lemberger_films}. Indeed, in these two-unit-cell thick
samples the superfluid-density downturn is necessarily between
$T_{BKT}^{n=1}$ and $T_{BKT}^{n=2}$, and its exact form depends on the
relative strength of $J_\perp$ and vortex-core energy. As we shall see,
this allows us to extract from the data of
Ref.~[\onlinecite{lemberger_films}] the doping dependence of the
vortex-core energy.

As a starting point we need a model for the BKT SC transition in the
two-layers system, where for simplicity we denote in the following with
``layer'' a single unit cell of the sample.  By adopting the formal analogy
between quantum 1D and thermal 2D
systems,\cite{benfatto_kt_rhos,giamarchi_book_1d} we describe each layer
(labeled with the subscript $1,2$, respectively) as a quantum 1D
sine-Gordon model ($\int\equiv \int dx\, v_s/(2\pi)$):
\bea
\lb{h0}
H_{1,2}=H^0_{1,2}-
\frac{g_u}{a^2}\int \cos(2\phi_{1,2}),\\
\lb{sg}
H^0=\int \left[K(\pd_x\theta_{1,2})^2+\frac{1}{K}
(\pd_x\phi_{1,2})^2\right],
\eea
Here the $\theta_i$ represent the SC phases, and $\phi_i$ are the conjugate
fields, with $[\theta_i(x'),\pd_x \phi_j(x)]=i\pi\d_{ij}\d(x'-x)$, $K$ is
the Luttinger-liquid parameter, $g_u$ is the strength of the sine-Gordon
potential, $a$ is the short-distance cut-off and $v_s$ the velocity of 1D
fermions (which is immaterial in the 1D-2D mapping where $v_s\tau$ plays
the role of the second spatial
dimension,\cite{benfatto_kt_rhos,giamarchi_book_1d} $\tau$ being the
imaginary time). A vortex configuration for the $\theta$ variable requires
that $\oint \nabla \theta =\pm 2\pi$ over a closed loop, i.e. the creation
of a $2\pi$ kink, generated by the exponential of its conjugate
field, the operator $e^{i\phi}$.\cite{giamarchi_book_1d} Thus, the
parameter $K$ defines the superfluid stiffness and $g_u$ the vortex
fugacity, as:
\be
\lb{defk} K\equiv \frac{\pi J}{T}, \quad g_u=2\pi e^{-\beta
\mu},
\ee
where $J\equiv J_{n=1}$ is the single-layer stiffness.
Within the standard 2D XY model for the phase the vortex-core energy $\mu$
is controlled by $J$ itself:\cite{kosterlitz_thouless,review_minnaghen}
\be
\lb{eqmu}
\mu_{XY}=\pi J \ln(2\sqrt{2}e^\gamma)\simeq \frac{\pi^2}{2} J,
\ee
where $\gamma$ is the Euler's constant. Even though we will treat $\mu$ as
an {\em independent} parameter to be fixed by comparison with the
experiments, for the sake of clarity we will measure it in multiples of
$\mu_{XY}$. The effect of the Josephson coupling $J_\perp$ is
accounted for by the term:
\be
\lb{perp}
H_\perp=-\frac{g_\perp}{ a^2}\int \cos(\theta_1-\theta_2),
\ee
where $g_\perp=\pi J_\perp/T$. As we shall see below, the interlayer
coupling is relevant under renormalization group (RG) flow and tends to
lock the phases in neighboring layers. When this effect dominates over the
vortex unbinding, the superfluid density is not affected by crossing
the $T^{n=1}_{BKT}$. As the $T$ increases further an additional coupling
generated under RG flow becomes relevant and induces the 2D transition at
$T_{BKT}^{n=2}$. It corresponds to the formation of a vortex simultaneously
in the two layers:
\be
\lb{as}
\frac{g_s}{a^2}\int \cos(2(\phi_1+\phi_2)).
\ee
It is then clear that a more convenient basis to study the system is given
by the symmetric/antisymmetric fields,
$\theta_{s,a}=(\theta_1\pm\theta_2)/\sqrt{2}$. The full Hamiltonian then
becomes 
\bea
H&=&H^0_a(K_a)+H^0_s(K_s)+\frac{4 g_u}{a^2}\int \cos(\sqrt 2
\phi_s)\cos(\sqrt 2 \phi_a) \nn\\
&-&\frac{2 g_\perp}{a^2}\int \cos(\sqrt 2\theta_a)
+\frac{2 g_s}{a^2}\int \cos(2\sqrt 2\phi_s),
\eea
where $H^0$ is defined in Eq.\ \pref{sg}, $K_{s,a}=K$ and the initial value
of $g_s$ is zero, even though it is generated at ${\cal O}(g_u^2)$ under RG
flow (see Eq.~\pref{eqas} below).  The superfluid density $J_s$ is
connected to the second-order derivative of the free energy with respect to
an infinitesimal twist $\d$ of the phase, $\pd_x\theta_i\ra
\pd_x\theta_i-\d$.  Since the $\theta_a$ field is unchanged by this
transformation while $\pd_x\theta_s\ra \pd_x\theta_s-\sqrt{2}\d$, we
immediately see that $J_s$ is given by the asymptotic value of $K_s(\ell)$
under RG flow:
\be
J_s\equiv\frac{\hbar^2\rho_s^{2d}}{4m}=\frac{K_s(\ell\ra \infty)T}{\pi}.
\ee
The perturbative RG equations for the couplings $K_a,K_s,g_u,g_\perp,g_s$
can be derived by means of the operator product expansion. The result is
(see also \cite{benfatto_kt_rhos,donohue_bosonic_ladders,cazalilla_bilayer,
mathey_bilayer}):
\bea
\lb{eqka}
\frac{dK_a}{d\ell}&=&2g_\perp^2-K_a^2g_u^2,\\
\lb{eqgu}
\frac{dg_u}{d\ell}&=&(2-\frac{K_a+K_s}{2})g_u-g_ug_sK_s,\\
\lb{eqks}
\frac{dK_s}{d\ell}&=&-g_u^2K_s^2-2g_s^2K_s^2,\\
\lb{eqgj}
\frac{dg_\perp}{d\ell}&=&\left(2-\frac{1}{2K_a}\right)g_\perp,\\
\lb{eqas}
\frac{dg_s}{d\ell}&=&(2-2K_s)g_s+\frac{1}{2}g_u^2(K_s-K_a),
\eea
with $\ell=\log(a/a_0)$, where $a_0,a$ are the original and RG rescaled
lattice spacing, respectively. Eqs.~\pref{eqka}-\pref{eqas} share many
similarities with the multi-layer case discussed in
Ref.~[\onlinecite{benfatto_kt_rhos}].  If the layers are uncoupled
($g_\perp=0$) then $K_s=K_a$ at all scales and the transition occurs when
$g_u$ flows to strong coupling. This happens at $K_s=2$, which corresponds
to $T_{BKT}^{n=1}$ according to the definitions \pref{defj},\pref{jump} and
\pref{defk}. However, when the layers are coupled $g_\perp$ grows under RG
flow, even if it has initially a small value. If the bare couplings -which
are $T$ dependent- are such that $g_\perp$ becomes of order one before than
$g_u$, the Josephson coupling term will lock the relative phase $\theta_a$
in neighboring layers, and $g_u$ will flow to zero even if
$K_s<2$. However, as soon as $K_s=2$ the $g_s$ coupling becomes relevant,
signaling the simultaneous vortex formation in the two layers. This effects
makes $g_u$ relevant as well and the superfluid stiffness $K_s$ jumps
suddenly from the value $K_s=1$ at $T_{BKT}^{n=2}$ to zero.
\begin{figure}[htb]
\includegraphics[scale=0.3,angle=-90]{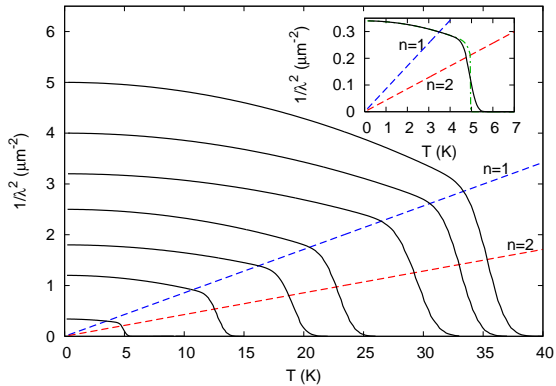}
\caption{(Color online) Temperature dependence of the superfluid density in the
  inhomogeneous bilayer system, with parameter values explained in the
  text. The downturn is located at the intersection with the line $n=1$ or
  $n=2$ for uncoupled or totally coupled layers, respectively.  Inset:
  expanded view for the most underdoped case.The dashed-dotted
  line is the result obtained without averaging over the $J_0$
  inhomogeneity.}
\label{fig-sup}
\end{figure}

As it was discussed in Ref.~[\onlinecite{benfatto_kt_rhos}], the range of
temperature above $T_{BKT}^{n=1}$ where the interlayer coupling allows the
system to sustain a finite superfluid density is {\em not} universal, and
depends crucially on the value of the vortex-core energy, which sets the
initial value of the fugacity $g_u$, see Eq.~\pref{defk}. Thus, at small
$\mu$ the system will display a rapid downturn already at $T_{BKT}^{n=1}$,
followed by an abrupt jump at $T_{BKT}^{n=2}$, while for large values of
$\mu$ we expect to see only the signature of the 2-layers BKT transition at
$T_{BKT}^{n=2}$. This is indeed what we observe in the experimental data of
Ref.~[\onlinecite{lemberger_films}]. While at higher dopings (25 K$<T_c<40$
K) the superfluid density shows a clear downturn already at
$T_{BKT}^{n=1}$, as the doping is decreased further this signature
disappears and only the BKT jump at $T_{BKT}^{n=2}$ is visible (see inset
of Fig.\ 1). To have a
quantitative estimate of $\mu$ we calculated $J_s$ by numerical integration
of the Eqs.~\pref{eqka}-\pref{eqas}, stopped at a scale $\ell^*$ where
$g_\perp=s\sim {\cal O}(1)$ (we used $s=3$), to account for the
perturbative character of the RG equations.  The bare temperature
dependence of the superfluid density mimics the low-$T$ behavior of the
data, $J(T)=J(T=0)-\a T^2$, with $J(T=0)$ and $\a$ extracted from the
experimental data well below the transition. We also assume that $J_\perp/J$
is independent on doping, and we choose $J_\perp/J=10^{-3}$ which is
appropriate in the range of doping considered\cite{hosseini_caxis}.
Thus, the only remaining free parameter is $\mu$, that can be chosen by
fitting the temperature dependence of the data around the transition. 
The result, reported in Fig.\ \ref{fig-sup} as $\l^{-2}(T)$, is in
excellent agreement with the data of Ref.\ \cite{lemberger_films}. 

\begin{figure}[htb]
\includegraphics[scale=0.3,angle=-90]{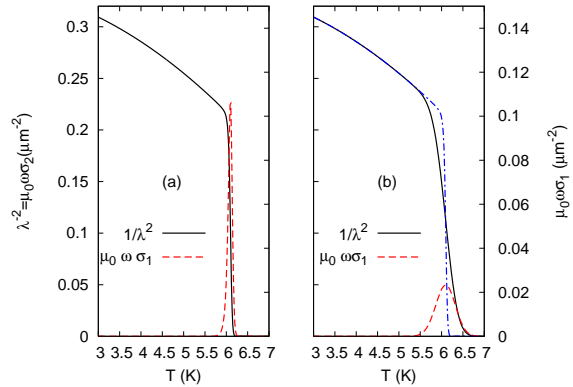}
\caption{(Color online) (a)
 $1/\l^2$ and $\m_0\o\s_1$ evaluated at $\o=50$ KHz for a single 
 $\bar J(T)$ curve (here $\mu=3\mu_{XY}$). The
 finite frequency leads to a sharp but continuous decrease of
 $1/\l^2$ across $T_{BKT}$, along with a peak in $\s_1$. (b) $1/\l^2$ and
 $\m_0\o\s_1$ evaluated at finite frequency using the averaged
 $J_{inh}$. It is also shown for comparison the homogeneous curve of panel
 (a) (dashed-dotted line).}
\label{fig-cond}
\end{figure}
According to the previous discussion, at $T_{BKT}^{n=2}$ the superfluid
density should display a BKT jump, while the data of
Ref.~\cite{lemberger_films} show clearly a broad tail around the estimated
$T_{BKT}^{n=2}$.  This effect, along with the temperature dependence of the
real part of the conductivity, cannot be attributed only to the finite
frequency of the measurements. Instead, the simplest explanation is the
existence of a $T_c$ inhomogeneity in the sample, which we accounted for in
the fit of Fig.\ 1. To clarify this point let us
discuss for simplicity the pure 2D case for a two-layers thick film. In
this case, after a transient regime the RG flow of the
Eqs.~\pref{eqka}-\pref{eqas} simplify to:
\be
\lb{eqgus}
\frac{dK_s}{d\ell}=-2K_s^2g_s^2, \quad
\frac{dg_s}{d\ell}=(2-2K_s)g_s,
\ee
with a fixed point at $K_s=1$, which corresponds to $T_{BKT}^{n=2}$ in
Eq.~\pref{jump}.  The complex conductivity $\s=\s_1+i\s_2$ at a finite
frequency $\o$ is given by:
\be
\lb{defs}
\s(\o)=-\frac{1}{\l^2e^2\m_0}\frac{1}{i\o \e(\o)},
\ee
where $\e(\o)=\e^1+i\e^2$ is complex dielectric constant, due to bound and
free vortex excitations. Following the dynamical theory of Ambegaokar et
al.\cite{ambegaokar_fin_freq} and Halperin and
Nelson\cite{halperin_ktfilms} we estimate these two contributions using the
RG flow \pref{eqgus} of the BKT coupling, and evaluating $\epsilon(\omega)$
at the finite scale $\ell_\o=\log(r_\o/a)$. Here $r_\omega=\sqrt{14 D/\o}$
is the maximum length probed by the oscillating field, where $D\sim
\hbar/m=10^{16}$ \AA$^2$/s is the diffusion constant of vortices and
$\omega$ the frequency of the measurements. According to Eq. \pref{defs}
and \pref{defj} $\l^{-2}=\mu_0 \omega \sigma_2$. Due to the finite
frequency, the jump of $\l^{-2}$ expected in the $\o=0$ case is replaced by
a sharp but continuous drop in a range $\D T_\o$ above $T_{BKT}^{n=2}$.  At
the same time $\s_1$ acquires a finite value, with a peak of approximately
the same width $\D T_\o$. However, using the value $\omega=50$ KHz
corresponding to the experiments of Ref.~\cite{lemberger_films} the
rounding effect on $\l^{-2}$ and the peak width in $\s_1$, reported in
Fig.\ \ref{fig-cond}a, are still much smaller than what measured
experimentally. A more reasonable explanation for the large transient
region is the sample inhomogeneity. Such inhomogeneity is also suggested by
tunneling measurements in other families of cuprates,\cite{yazdani_07}
where approximately Gaussian fluctuations of the local gap value are
observed. Even though the issue of the microscopic origin of this effect is
beyond the scope or this paper, nonetheless we find that an analogous
distribution of the superfluid-stiffness $J_0$ values around a given $\bar
J$ can account very well for the data of
Ref.~[\onlinecite{lemberger_films}]. Thus, we compare with the experiments
the quantity $J_{inh}(T)=\int dJ_0 P(J_0)J(T,J_0)$, where each $J(T,J_0)$
curve is obtained from the RG equations \pref{eqka}-\pref{eqas} using a
bare superfluid stiffness $J=J_0-\a T^2$. Each initial value $J_0$ has a
probability $P(J_0)=\exp[-(J_0-\bar J_0)^2/2\s^2](\sqrt{2\pi} \s)$ of being
realized, where the bare average stiffness $\bar J(T)=\bar J_0-\a T^2$ has
$\bar J_0$ and $\a$ fixed by the experimental data at low $T$, where
$J_{exp}(T)$ is practically the same as $\bar J(T)$. However, using a
variance $\s=0.05 \bar J_0$ we obtain a very good agreement with the
experiments near the transition, as far as both both the tail of $\l^{-2}$
and the position and width of $\s^1(\omega)$ are concerned, see Fig.\
\ref{fig-cond}b.  Observe also that such a variance can be compatible,
within an intermediate-coupling scheme for the superconductivity, with the
few times larger distribution of gap values ($\sigma\approx 0.15
\bar\Delta$) reported in tunneling experiments.\cite{yazdani_07}

\begin{figure}[htb]
\includegraphics[scale=0.3,angle=-90]{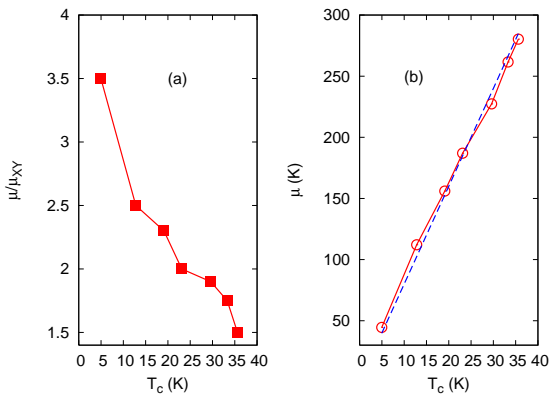}
\caption{(Color online) Vortex-core energy as a function of $T_c$ extracted
 from the fit in Fig.\ \ref{fig-sup}. (a) Ratio between $\mu$ and the
 $\mu_{XY}$ value \pref{eqmu}, proportional to the single-layer
 energy $J$. (b) Absolute value of $\mu$ in K. The dashed line is $\mu=8 T_c$.}
\label{fig-mu}
\end{figure}

The same finite-frequency analysis is made more involved in the bilayer
case, because the RG equations \pref{eqka}-\pref{eqas} should be stopped at
scales smaller than $\ell_\o$, to prevent the flow of the $g_\perp$ at
strong coupling. Thus, in Fig.\
\ref{fig-sup} we included only the effect of the $T_c$ inhomogeneity by
means of the average over $P(J_0)$ discussed above (with $\sigma=0.06 \bar J_0$
for the most underdoped samples).  This procedure accounts
very well for the long tails of $\l^{-2}$ above the transition (see inset),
without affecting significantly the estimate of $\mu$.

Let us now comment on the doping dependence of $\mu(T=0)$ reported in Fig.\
\ref{fig-mu}. As we said, the measured $T_c$ crosses over from
approximately $T_{BKT}^{n=1}$ at high doping to $T_{BKT}^{n=2}$ at low
doping. This is reflected in the doping dependence of $\mu/\mu_{XY}$:
indeed, as we observed for the multi-layer case,\cite{benfatto_kt_rhos} as
$\mu$ increases with respect to the single-layer stiffness $J$, the
transition moves away from $T_{BKT}^{n=1}$.  It is worth noting that since
$T_c$ is controlled by the competition between the vortex fugacity and the
interlayer coupling, in principle the crossover of $T_c$ from
$T_{BKT}^{n=1}$ to $T_{BKT}^{n=2}$ could be obtained also by keeping $\mu$
fixed and by varying $J_\perp$. However, to reproduce the data we should
assume in this case an unlikely increase of $J_\perp$ by two order of
magnitudes as the doping is decreased, up to $J_\perp/J\sim 10^{-1}$.
Despite the non-universal behavior of $\mu/\mu_{XY}$, the absolute value of
$\mu$ reported in Fig.\ \ref{fig-mu}b scales linearly with $T_c$.  This is
our central result, which establishes a precise
relation between the vortex-core energy and $T_c$ in severely underdoped
cuprate superconductors.

%we would like to comment on a possible
%interpretation of it within the strong-coupling limit of the attractive
%Hubbard model. Let us assume that the BCS-like relation $\mu\simeq E_c
%\xi_0^2$ is still valid in this regime, where $\xi_0$ is also a measure of
%the correlation length of Cooper pairs, which is always of order of the
%lattice spacing in the strong-coupling
%limit\cite{benfatto_coherence_length}. At the same time, recent DMFT
%calculations on the Hubbard model have shown that at strong coupling the
%condensation-energy density scales with $T_c$\cite{toschi_attractiveU}. In
%this picture the scaling of $\mu$ with $T_c$ is then a signal that the
%system is entering a strong-coupling regime for superconductivity, where
%indeed a large separation between $\D$ and $T_c$ is expected, $T_c$ being
%controlled by phase fluctuations only.

In summary, we analyzed the occurrence of the BKT transition in a bilayer
system. By means of RG approach we computed the temperature dependence of
the superfluid stiffness of the bilayer, and we proved the crucial role
played by the vortex-core energy $\mu$ in controlling the
transition. Taking into account also the sample inhomogeneity we provided
an excellent fit of the experimental data of
Ref.~[\onlinecite{lemberger_films}], which allowed us to extract for the
first time a linear scaling of $\mu$ with $T_c$ in underdoped YBCO.  A
theoretical understanding of this result is still lacking and no doubt it
would constitute a stringent test of microscopic proposals for the
underdoped phase.

%Finally, notice that
%for strong underdoped samples the large chemical-potential value would lead
%to a negligible renormalization of $J_s$ with respect to $J$ below
%$T_{BKT}^{n=2}$ if the 2-unit-cell film is treated as a single 2D system,
%see Fig.\ 3a.  Instead, in the calculation with a layered model reported in
%Fig.\ 2 a partial renormalization of $J_s$ below $T_{BKT}^{n=2}$ is an
%effect reminiscent of the single-layer vortices, effective already above
%$T_{BKT}^{n=1}$.
This work was supported in part by Miur PRIN 2005, Prot. 2005022492, and by
the Swiss NSF under MaNEP and division II.

\end{document}